\documentclass[final,5p,times,twocolumn]{elsarticle}
\usepackage{amsmath,amssymb}
\usepackage{color}
\usepackage{graphicx}
\usepackage{tabularx}
\usepackage{lineno}
\modulolinenumbers[5]

\usepackage[euler]{textgreek}
\usepackage{siunitx}
\DeclareSIUnit{\wtpercent}{wt.\%} 
\DeclareSIUnit{\atpercent}{at.\%} 
\usepackage{miller}
\usepackage{textcomp}
\usepackage{units}

\journal{Acta Mater.}

\begin{document}

\begin{frontmatter}

\title{Mechanisms of Ti$_3$Al precipitation in hcp $\alpha$-Ti}

\author[IC]{Felicity F. Dear}
\author[MPIE]{Paraskevas Kontis}
\author[MPIE,IC]{Baptiste Gault}
\author[APS]{Jan Ilavsky}
\author[IC,RR]{David Rugg}
\author[IC]{David Dye}

\address[IC]{Department of Materials, Royal School of Mines, Imperial College London, Prince Consort Road, London, SW7 2BP, UK}
\address[MPIE]{Max-Planck Institut f\"{u}r Eisenforschung GmbH, Max-Planck-Stra\ss{}e 1, D\"{u}sseldorf, Germany}
\address[APS]{Argonne National Laboratory, 9700 S. Cass Avenue, Building 433A, Argonne, IL 60439, USA}
\address[RR]{Rolls-Royce plc, Elton Road, Derby, DE24 8BJ, UK}

\begin{abstract}\begin{small}
Nucleation and growth of Ti$_3$Al \textalpha{}$_2$ ordered domains in \textalpha{}-Ti--Al--X alloys were characterised using a combination of transmission electron microscopy, atom probe tomography and small angle X-ray scattering. Model alloys based on Ti--7Al~(wt.\%) and containing O, V and Mo were aged at \SI{550}{\celsius} for times up to \SI{120}{\day} and the resulting precipitate dispersions were observed at intermediate points. Precipitates grew to around \SI{30}{\nano\metre} in size, with a volume fraction of 6--10\% depending on tertiary solutes. Interstitial O was found to increase the equilibrium volume fraction of \textalpha{}$_2$, while V and Mo showed relatively little influence. Addition of any of the solutes in this study, but most prominently Mo, was found to increase nucleation density and decrease precipitate size and possibly coarsening rate. Coarsening can be described by the Lifshitz-Slyozov-Wagner model, suggesting a matrix diffusion-controlled coarsening mechanism (rather than control by interfacial coherency). Solutionising temperature was found to affect nucleation number density with an activation energy of $E_{\mathrm{f}} = 1.5\pm{}0.4$~eV, supporting the hypothesis that vacancy concentration affects \textalpha{}$_2$ nucleation. The observation that all solutes increase nucleation number density is also consistent with a vacancy-controlled nucleation mechanism.
\end{small}\end{abstract}


\end{frontmatter}

\section{Introduction}
\label{Intro}

Predicting the lifetime of safety-critical components in gas turbine engines is crucial to continued improvements in flight safety. Despite a global increase of 90\% in passenger journeys and 40\% in freight transport by air in the decade to 2018 \cite{WB}, the accident rate across the same time period dropped by 30\% \cite{ICAO}. Sustaining this safety improvement relies on mechanistic understanding of the materials serving in such applications, and this same knowledge can steer the development of more capable alloys.

Titanium alloys form a key materials system for aerospace applications, with \textalpha{}+\textbeta{} alloys such as Ti--64 (Ti--6Al--4V, wt.\%) widely used in fan and compressor components \cite{Boyer1996}. Deeper understanding of these alloys' response to fatigue loading has been an area of intense effort for several years \cite{Evans1994,Brandes2010}.

In \textalpha{}+\textbeta{} alloys, the elastic anisotropy and limited slip system availability of the hcp \textalpha{} phase have a significant impact on the polycrystalline material's overall response to static and cyclic loading regimes \cite{Dunne2007}. This means that, during high cycle or dwell fatigue loading, variations in elastic and plastic behaviour from grain to grain can initiate yielding near the boundary of particularly mismatched grains.

The nature of this intragranular plasticity plays a key role in fatigue crack initiation. If easy cross-slip is possible, dislocations are able to travel across an individual grain to intersect its grain boundaries at any location. If dislocations are instead restricted in their ability to cross-slip and travel homogeneously, slip bands are formed and eventually intersect a grain boundary. This results either in slip transmission, where the next grain in its path is well oriented for deformation, or in a dislocation pile-up \cite{Joseph2018}. A sufficiently large pile-up may impose enough stress to nucleate a fatigue crack \cite{Dunne2008}. Groups of similarly-oriented grains (macrozones) and slip bands extending across millimetres have been implicated in dwell fatigue \cite{PilchakFAA,Evans1994,Neal1988}, including in service issues \cite{BEAReport}.
 Slip band formation and factors promoting the even distribution of slip across the \textalpha{} microstructure are hence of significant interest.

Aluminium is commonly included at around \SI{6}{\wtpercent} in \textalpha{}+\textbeta{} alloys, stabilising the \textalpha{} phase and providing solid solution strengthening. Phase segregation produces an \textalpha{} composition closer to \SI{7}{\wtpercent} (\SI{11.8}{\atpercent}) meaning that, at temperatures of 500--\SI{700}{\celsius}, crystallographic ordering of Al can occur and lead to precipitation of the \textalpha{}$_2$ phase (Ti$_3$Al, DO$_{19}$ structure) \cite{Gehlin1970}.
The position of the \textalpha{}/\textalpha{}+\textalpha{}$_2$ boundary has been the subject of several studies, with successive iterations of the Ti--Al phase diagram shifting it towards lower Al content \cite{Namboodhiri1973,Sircar1986}. A region of short-range ordering (SRO) between truly disordered \textalpha{} and the \textalpha{}+\textalpha{}$_2$ field has also been proposed \cite{Namboodhiri1983}. Recent diagrams place the boundary at around \SI{10}{\atpercent} in calculated diagrams and \SI{12}{\atpercent} in Schuster and Palm's diagram drawing together numerous experimental observations \cite{Witusiewicz2008,Wang2012,Schuster2006}.

Even in the early stages of \textalpha{}$_2$ formation, where some form of ordering (evidenced by faint superlattice reflections) exists prior to precipitates being resolvable in dark field TEM images, there is a significant impact on slip behaviour \cite{Neeraj2001}. Cross-slip is hindered, such that first dislocations passing across a grain disrupt the local ordering, leaving a trail of disrupted structure that offers an easier route for subsequent dislocations. This results in slip band formation and the associated deleterious micromechanical effects, with notable implications for tensile and fatigue response \cite{Brandes2010}. The earliest stages of ordered domain formation have been shown to restrict primary creep in \textalpha{}-Ti \cite{Neeraj2000,Neeraj2001}. Dislocation pinning by \textalpha{}$_2$ precipitates has also been observed \cite{Williams1972}. In macroscopic plastic deformation, \textalpha{}$_2$ causes initial strain hardening followed by localised strain softening\cite{Neeraj2001,Williams1969}, due to initial resistance to slip by the ordered domains being overtaken by the establishment of slip bands as easy paths for slip.

Studies on the formation mechanism of \textalpha{}$_2$ are challenging due to the nanometre length scales and small compositional variations between matrix and precipitate that are involved, especially during the early stages of phase separation. Long-term ageing at temperatures around 500--\SI{600}{\celsius} typically produces precipitates 5--\SI{10}{\nano\metre} in size. Morphology is spheroidal, with elongation along the shared \textit{c}-axis occurring as growth proceeds and the precipitates undergo coarsening \cite{Blackburn1967}. The earliest stages of phase separation remain less well understood. Possible mechanisms include homogeneous nucleation or spinodal decomposition, and a spinodal decomposition triggered by SRO has also been suggested \cite{Blackburn1967,Liew1999,Wood1998}.

The influence of additional alloying elements on \textalpha{}$_2$ formation is a significant consideration in trying to gauge the propensity of commercial alloys to deleterious Al ordering. Interstitial oxygen content has been shown to promote \textalpha{}$_2$ formation, shifting the \textalpha{}/\textalpha{}+\textalpha{}$_2$ phase boundary to lower Al content and higher temperature \cite{Schuster2006,Waterstrat1988,Lim1976,Gray1990,Bagot2018}. This is suggested to be caused by a reduction in the solubility of Al in Ti with increasing O content \cite{Lim1976}. The presence of \textbeta{} stabilisers in \textalpha{}+\textbeta{} alloys is also thought to influence \textalpha{}$_2$ formation \cite{Radecka2016}.

This work investigates the factors controlling \textalpha{}$_2$ formation in an isothermal ageing study of a model alloy series based on Ti--7Al (wt.\%), with additions of O, V and Mo. Microstructures were observed in TEM, and local compositions were measured in atom probe tomography. Precipitate dispersion parameters such as number density were analysed using small angle X-ray scattering. Insights are then drawn regarding the role of vacancies in nucleation, a hypothesis which is then tested, allowing the role of solutes in enhancing \textalpha{}$_2$ precipitation to be understood.

\section{Experimental methodology}
\label{experimental}

Alloys listed in Table~1 were melted from Ti sponge (Toho, Japan), TiO powder and pure Al, V and Mo pellet in an Arcast200 \SI{27}{\kilo\watt} low pressure argon arc melter and cast to produce 23$\times$23$\times$\SI{55}{\milli\metre} ingots. The alloys were then rolled and recrystallised in the \textalpha{} phase followed by ice water quenching. Samples of each alloy in this IWQ (disordered) condition were taken, and \SI{10}{\milli\meter} cubes were then encapsulated under an Ar atmosphere in quartz and aged at \SI{550}{\celsius} for up to \SI{120}{\day} and furnace cooled, to evolve the \textalpha{}$_2$ precipitate dispersions. Separately, samples of each material from the quenched condition were subjected to 2~hours ageing at \SI{550}{\celsius} before air cooling (AC condition), to capture the early stages of ordering.

Microstructures were initially observed in backscatter electron imaging on a Zeiss Sigma~300 FEG-SEM operated at \SI{8}{\kilo\volt}. SEM specimens were prepared by electropolishing with a 3\% perchloric acid solution at \SI{-35}{\celsius} and \SI{20}{\volt}. Bulk compositions were confirmed using ICP-OES and combustion analysis provided by TIMET UK Ltd, Table~1.

\begin{table*}[h]\begin{small}
	\centering
	\caption{Compositions of the Ti--7Al model alloy series measured by ICP-OES and combustion analysis by TIMET (Witton, UK). The hydrogen content in each alloy was measured to be \SI{0.01}{\wtpercent} or less. Alloys in this study are referred to by their nominal compositions. Rolling ($T_{\mathrm{roll}}$) and recrystallisation ($T_{\mathrm{RX}}$) temperatures for these steps were chosen according to the \textbeta{} transus for each alloy, identified by iterative heat treatments and metallography. Recrystallisation times ($t_{\mathrm{RX}}$) were chosen to account for the varying recrystallisation kinetics in each system.}
	\begin{tabular}{l | c c c c c | c c c c c | c c c} 
	\hline
	Alloy (nominal & \multicolumn{5}{c|}{Measured composition / wt.\%} & \multicolumn{5}{c|}{Measured composition / at.\%} & $T_{\mathrm{roll}}$ & $T_{\mathrm{RX}}$ & $t_{\mathrm{RX}}$\\
	composition) & Al & V & Mo & O & N & Al & V & Mo & O & N & /\si{\celsius} &  /\si{\celsius} &  /h\\
	\hline
	Ti--7Al & 6.58 & $<$0.01 & $<$0.01 & 0.05 & 0.02 & 11.09 & $<$0.01 & $<$0.01 & 0.14 & 0.06 & 900 & 980 & 1 \\
	~--0.25O & 7.14 & $<$0.01 & $<$0.01 & 0.24 & 0.05 & 11.95 & $<$0.01 & $<$0.01 & 0.68 & 0.16 & 900 & 980 & 1 \\
	~--1.1V & 7.01 & 1.21 & $<$0.01 & 0.07 & 0.04 & 11.79 & 1.08 & $<$0.01 & 0.20 & 0.13 & 900 & 850 & 1 \\
	~--1.1V--0.25O & 7.04 & 1.17 & $<$0.01 & 0.26 & 0.08 & 11.80 & 1.04 & $<$0.01 & 0.73 & 0.26 & 900 & 850 & 1 \\
	~--0.8Mo & 7.17 & $<$0.01 & 0.83 & 0.10 & 0.08 & 12.08 & $<$0.01 & 0.39 & 0.28 & 0.26 & 850 & 850 & 18 \\
	~--0.8Mo--0.25O & 7.16 & $<$0.01 & 0.90 & 0.30 & 0.01 & 12.01 & $<$0.01 & 0.42 & 0.85 & 0.03 & 850 & 850 & 18 \\
	\hline
	\end{tabular}
	\label{table:process}
\end{small}\end{table*}

Formation of \textalpha{}$_2$ was observed with conventional TEM methods, with specimens prepared by jet electropolishing with a 3\% perchloric acid solution at \SI{-35}{\celsius} and \SI{20}{\volt} to perforation. Using a JEOL 2100F TEM operated at \SI{200}{\kilo\volt}, selected area electron diffraction patterns were collected for each sample along \hkl<0 1 -1 1> directions. Dark field images of the \textalpha{}$_2$ precipitates were then made using the \hkl{2 -1 -1 0}$_{\alpha_{2}}$ reflections. This provided qualitative observation of precipitate and dispersion characteristics, and allowed measurement of precipitate aspect ratios by measuring several wholly contained, clearly visible precipitates in images for each sample.

Measurement of compositional features of the precipitates at sub-nanometre resolution was performed using atom probe tomography (APT). Specimens were prepared by conventional Ga$^+$ FIB lift-out methods \cite{Thompson2007}. Specific grain orientations were targeted using EBSD mapping prior to FIB work. This produced specimens with the APT analysis direction oriented within a few degrees of the \hkl<2 -1 -1 0> zone axis. Titanium and Ti-based alloys are prone to forming hydrides during specimen preparation for TEM and APT, an artefact that can be avoided by performing the final thinning or sharpening of specimens at cryogenic temperatures \cite{Chang2019}. Here, we used the infrastructure described in \cite{Stephenson2018,Rivas2020} for cryogenic preparation, yet upon comparing with specimens obtained on the same FIB at room temperature, no significant differences in the H uptake were noticed. The low solubility of H in \textalpha{}-Ti and the targeted preparation far from any interfaces likely explains these observations. APT samples were then run on a Cameca LEAP 5000 XS operated in voltage mode at \SI{50}{\kelvin}, with a pulse frequency of 200--\SI{250}{\kilo\hertz}, pulse fraction of 20\% and detection rate of 0.20--0.40\%. The data collected were then analysed using Cameca's IVAS analysis suite for reconstruction and MATLAB scripts for further analysis.

Quantification of dispersion statistics and precipitate evolution was achieved using small angle X-ray scattering (SAXS). Specimens were prepared by electrodischarge machining of a \SI{3}{\milli\meter} diameter cylinder, from which discs of \SI{300}{\micro\metre} thickness were cut with a precision saw. The discs were then ground to the appropriate thickness ($\sim$\SI{100}{\micro\metre}) by hand using SiC grit papers up to a 4000 grit finish, followed by polishing with a neutralised colloidal silica solution by hand. After thorough cleaning with detergent to remove colloidal silica and isopropanol to remove surface contaminants, specimens were suspended in an amorphous, transparent tape for handling during measurement. SAXS measurements were then taken on the USAXS beamline \cite{Ilavsky2018} at the Advanced Photon Source at Argonne National Laboratory, using a \SI{21}{\kilo\electronvolt} beam and a 800$\times$\SI{800}{\micro\metre} beam area. The data were reduced using the Nika package \cite{Ilavsky2012} and analysed using the Irena package \cite{Ilavsky2009} within Igor Pro. Morphological information from TEM imaging and compositional information from APT were used to guide the fitting of SAXS data with known shapes, aspect ratios and phase compositions, allowing deconvolution of volume fraction and contrast.

\section{Results}
\label{results}

\subsection{Microstructural characterisation}

\begin{figure}[t!]
   \centering
   \includegraphics[width=90mm]{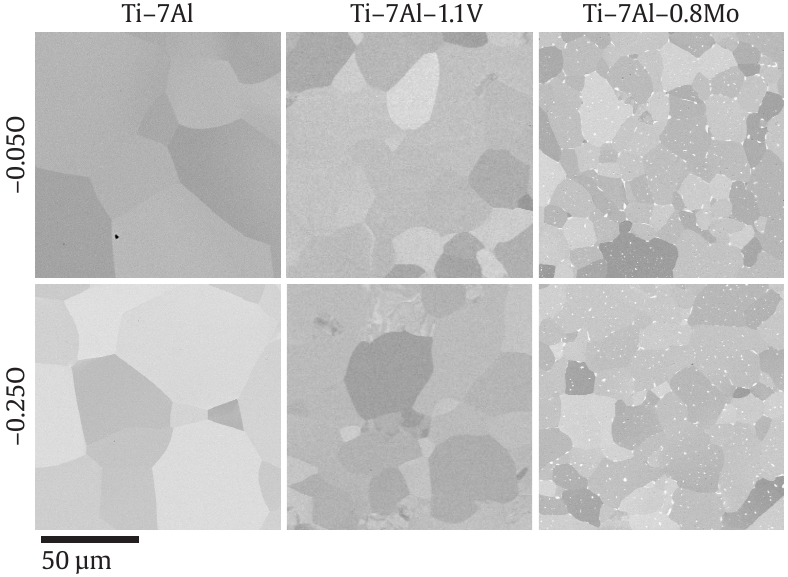}
   \caption{Backscatter micrographs of alloy microstructures, showing the intended equiaxed \textalpha{} microstructure. In the Mo-containing alloys, the very limited solubility of Mo in \textalpha{} led to the formation of small, micron-scale \textbeta{} domains at grain boundary triple points, which also contributed to grain size refinement during processing.}
   \label{fig:backscatter}
\end{figure}

Equiaxed \textalpha{} microstructures were produced, Fig.~\ref{fig:backscatter}, with grain sizes of 10--\SI{50}{\micro\metre} depending on alloy composition. Alloys containing no \textbeta{} stabilisers showed a larger grain size due to the limited opportunity to restrict grain size during processing of material with a very narrow \textalpha{}+\textbeta{} phase field. The Mo-containing alloys contained a small fraction of \textbeta{} due to the very low solubility for Mo in \textalpha{}; it was later demonstrated in APT results that the \textalpha{} phase contained a small amount of Mo, as intended.

\subsection{Transmission electron microscopy}

\begin{figure}[h!]
   \centering
   \includegraphics[width=80mm]{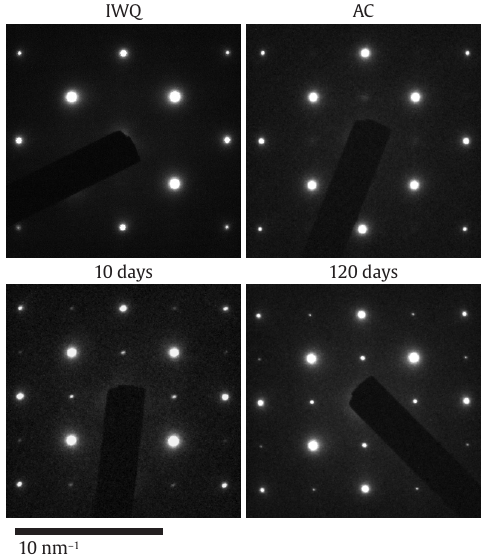}
   \caption{Selected area electron diffraction patterns (\textbf{B}~=~\hkl<0 1 -1 1>) obtained for Ti--7Al--0.05O~(wt.\%), in the ice water quenched (disordered) state, and in selected subsequent ageing states. Diffuse superlattice reflections are faintly visible after 2~hours ageing at \SI{550}{\celsius} (AC condition), which intensify as phase separation progresses at 10 days and 120 days.}
   \label{fig:saedp}
\end{figure}

Selected area electron diffraction patterns taken for \textbf{B}~=~\hkl<0 1 -1 1>, Fig.~\ref{fig:saedp}, show the development of superlattice reflections as the ordered \textalpha{}$_2$ phase forms and grows. After a short hold of 2~hours at \SI{550}{\celsius}, a small amount of intensity was observable at superlattice spot locations. Upon further ageing for 10~days or more, superlattice reflections became distinct spots and increased in intensity as ageing progressed.

\begin{figure*}[p]
   \centering
   \includegraphics[width=0.9\textwidth]{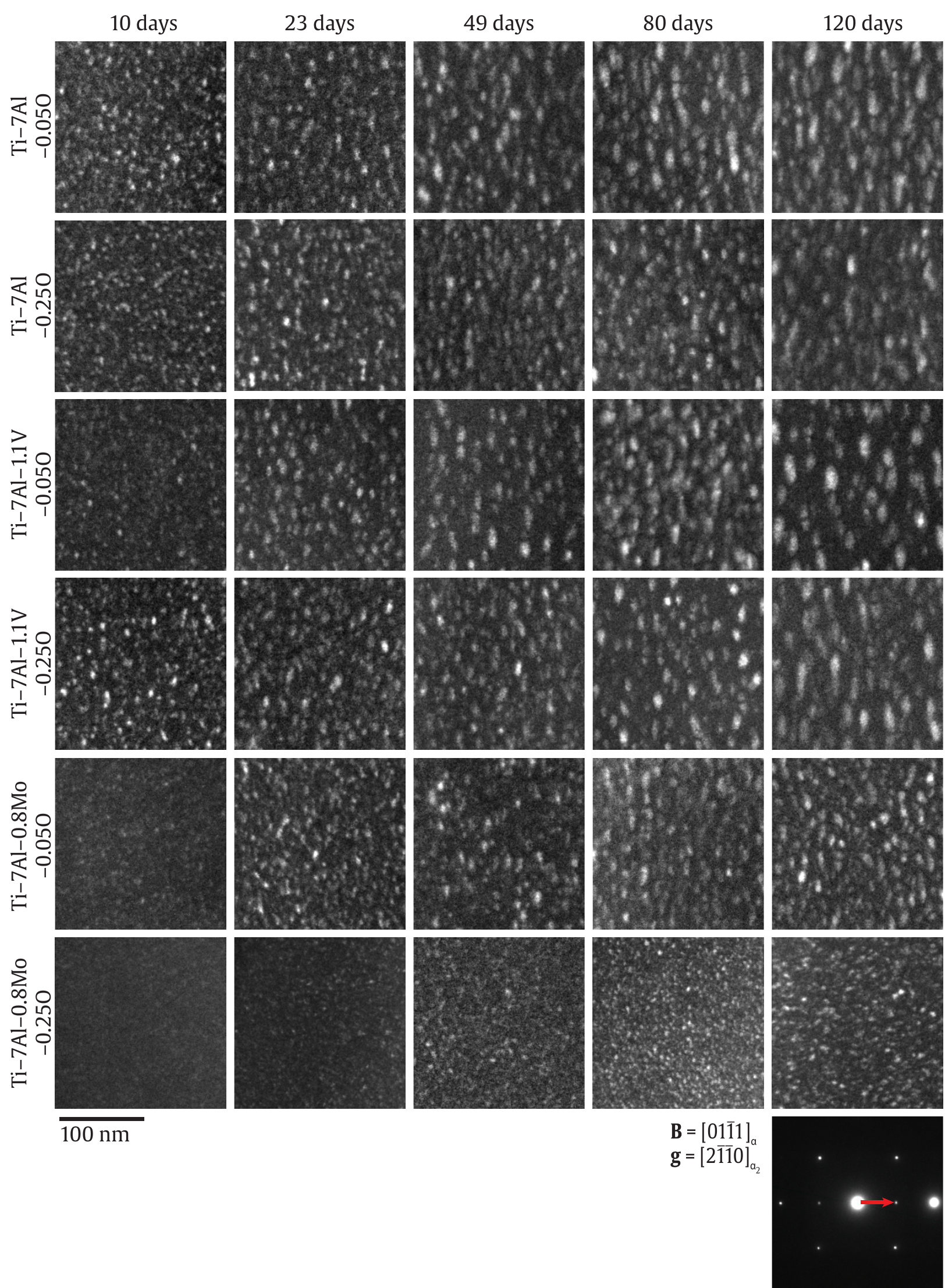}
   \caption{Dark field transmission electron micrographs recorded for specimens of each alloy at different ageing times, using a two-beam condition with the \hkl[2 -1 -1 0] reflection for \textbf{B}~=~\hkl[0 1 -1 1]. The base alloy Ti--7Al--0.05O~(wt.\%) shows formation of spheroidal precipitates that increase in size and aspect ratio as ageing progresses. Additional solutes modify the way in which precipitate size, aspect ratio, spacing and number density evolve over time.}
   \label{fig:darkfield}
\end{figure*}

Dark field images provide a qualitative view of trends in precipitate morphology, size and number density during ageing, Fig.~\ref{fig:darkfield}. Imaging was attempted for the 2~hour aged specimens, but no image contrast was evident. The base alloy Ti--7Al--0.05O showed formation of nanoscale precipitates after 10~days, which coarsened over time whilst growing in size and increasing in aspect ratio.

The addition of oxygen to the alloy system causes an increase in precipitate number density, and produces smaller precipitates. The effect of oxygen on volume fraction of the \textalpha{}$_2$ phase is unclear from qualitative micrographs; it should be recalled that these are projections of contrast through the foil thickness. TEM images do not show evidence of a significant effect of vanadium on the precipitate dispersion. The addition of molybdenum considerably restricts precipitate sizes, and precipitate aspect ratio does not increase as significantly over the duration of the study. However, these micrographs only provide a qualitative impression of the alloying trends; for a quantitative comparison we turn next to atom probe tomography and SAXS.

\subsection{Atom probe tomography}

Atom probe tomography results provided a quantitative analysis of local compositional features. Of specific interest were the compositions of phases present, segregation of solutes between these phases, compositional features of the \textalpha{}/\textalpha{}$_2$ interface, and the crystallographic site partitioning of V and Mo on the \textalpha{}$_2$ DO$_{19}$ lattice. Measurements were performed for Ti--7Al--0.05O in the quenched condition as a reference dataset, for this alloy aged for 49~days, and for this alloy and Ti--7Al--0.25O, Ti--7Al--1.1V--0.25O and Ti--7Al--0.8Mo--0.25O in the 120-day aged condition to observe \textalpha{}$_2$ precipitates.

Data reconstruction was informed by TEM observations of precipitate morphology and by crystallographic information about the material. Since specimens were prepared from a known crystallographic orientation, partial indexing of desorption maps was possible and confirmed that the pole approximately parallel to the analysis direction was \hkl<2 -1 -1 0> in each specimen. This allowed calibration of reconstruction parameters by guiding the reconstruction according to the known interplanar spacing of \hkl{2 -1 -1 0} planes in \textalpha{}-Ti.

\begin{figure}[t!]
   \centering
   \includegraphics[width=88mm]{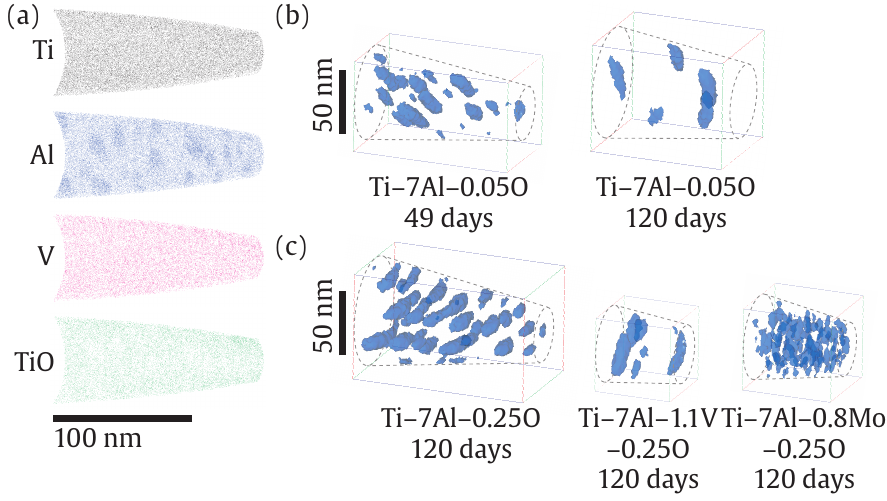}
   \caption{Examples of atom probe tomography results. (a) Atom maps showing the distribution of ion types detected for a specimen of Ti--7Al--1.1V--0.25O aged for 120 days at \SI{550}{\celsius}. Note the domains of increased Al content, which are indicative of \textalpha{}$_2$ phase formation. O was detected only within TiO$^{n+}$ complex ions. 8.5\% Al concentration isosurfaces were used to indicate phase boundaries in aged specimens. The coarsening of \textalpha{}$_2$ precipitates is evident in comparing Ti--7Al--0.05O after ageing at \SI{550}{\celsius} for 49 days and 120 days, (b). The precipitate dispersion characteristics seen in dark field TEM (Fig.~\ref{fig:darkfield}) are reflected in the Al isosurfaces shown for different alloys after ageing for 120 days, (c).}
   \label{fig:atmapiso}
\end{figure}

Analysis of Ti--7Al--0.05O in the quenched condition revealed a homogeneous distribution of all solutes in species density maps, with no indications of phase separation. It is noted that the presence of short-range ordering versus true disorder cannot necessarily be inferred from APT data due to sub-100\% ion detection efficiency. Upon ageing to 49 days, a dispersion of \textalpha{}$_2$ was evident in the specimen as regions of increased Al content, displaying the ellipsoidal morphology observed in TEM, Fig.~\ref{fig:atmapiso}. After 120 days at \SI{550}{\celsius}, this alloy displayed the expected coarsening of precipitates. APT observations of alloys containing O, V and Mo additions showed precipitate dispersions with characteristics as expected from earlier TEM observations, with increased number density upon adding these solutes and reduced precipitate size in the Mo-containing alloys.

\begin{figure}[t!]
   \centering
   \includegraphics[width=80mm]{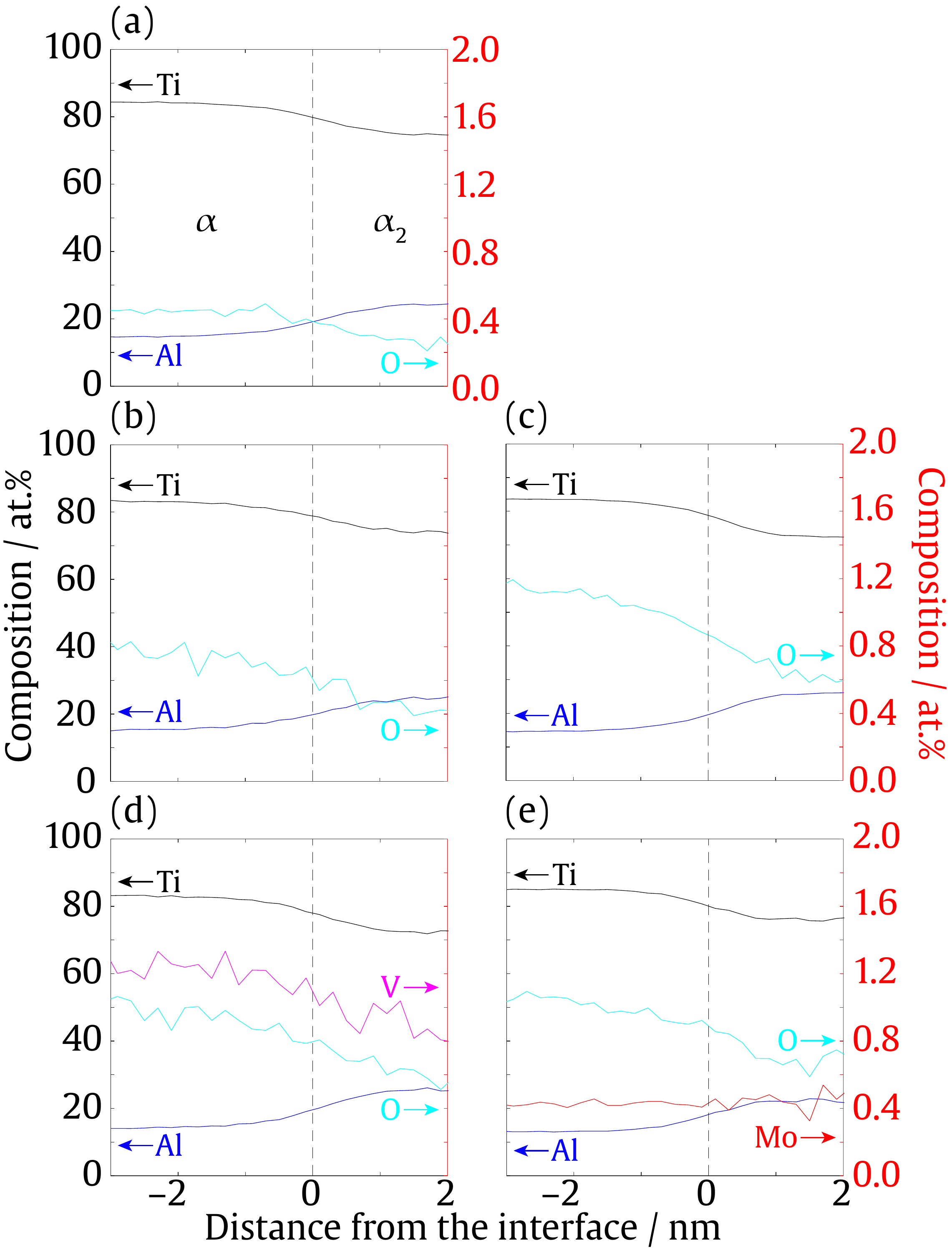}
   \caption{Proximity histograms constructed for the \textalpha{}/\textalpha{}$_2$ interface in APT datasets: (a) Ti--7Al--0.05O, aged 49~d; (b) Ti--7Al--0.05O, aged 120~d; (c) Ti--7Al--0.25O, aged 120~d; (d) Ti--7Al--1.1V--0.25O, aged 120~d; (e) Ti--7Al--0.8Mo--0.25O, aged 120~d. Segregation of aluminium to the \textalpha{}$_2$ at a concentration of approximately \SI{25}{\atpercent} is evident, corresponding well with the expected Ti$_3$Al stoichiometry. Notably, oxygen is found to segregate to the matrix \textalpha{} phase in each alloy, despite the fact that it promotes \textalpha{}$_2$ formation. Vanadium shows a similar behaviour to oxygen, while molybdenum showed no clear segregation to either phase. (Ti, Al shown against black axis; O, V, Mo shown against red axis.)}
   \label{fig:proxi}
\end{figure}

Proximity histograms (composition profiles calculated as a function of the distance to a specific isosurface \cite{Hellman2000}) were produced to analyse the nature and extent of phase segregation for each solute, Fig.~\ref{fig:proxi}. The \textalpha{}$_2$ phase was identified according to its Al enrichment to around \SI{25}{\atpercent}.

Elements seen to promote \textalpha{}$_2$ formation in dark field TEM observations were expected to show segregation to this phase. For both O and V, segregation to the \textalpha{} matrix was instead observed. Previously we have shown \cite{Bagot2018} that O enhances \textalpha{}$_2$ formation whilst segregating to the $\alpha$ phase, owing the curvature of the phase boundary in the Ti--Al--O ternary system. Mo showed no segregation between the phases despite its significant effect on the \textalpha{}$_2$ precipitate dispersions.

For each specimen, proximity histograms were used to choose values for a set of Al concentration isosurfaces at 6.5\% to select \textalpha{} and 10.5\% to select \textalpha{}$_2$ without including the interfacial region. This approach was used to obtain phase compositions, Table~\ref{table:aptcomp}. These phase compositions were used to provide contrast values for SAXS analysis, allowing deconvolution of the volume fraction and compositional contributions to peak size in the SAXS data.

The analysed \textalpha{} compositions appear low in Ti (or, equivalently, high in Al) compared to the bulk ICP-OES analysis (Table~\ref{table:process}). This is a direct and unavoidable consequence of the large difference in evaporation field between Ti and Al, causing Ti to be preferentially under-counted due to multiple evaporation events \cite{Kingham1982,Peng2018}. This bias is more pronounced in the \textalpha{} phase, and it should thus be noted that this could cause a slight underestimate (on the order of a few per cent) of SAXS contrast and hence a slight overestimate of the volume fractions presented in the below analysis of SAXS data. This consideration was incorporated in estimates of uncertainty for the dispersion characteristics calculated from the SAXS data.

\begin{table}[h!]\begin{small}
	\centering \setlength{\tabcolsep}{4pt}
	\caption{Bulk and phase compositions measured in APT (no background correction applied). Bulk compositions were analysed counting all ions in a dataset. For phase compositions, proximity histograms were used to determine values for a set of aluminium isoconcentration surfaces to isolate each phase, excluding the interfacial region.}
	\label{table:aptcomp}
	\begin{tabular}{l l l r r r r r r} 
	\hline
	Material & Ageing &  & \multicolumn{6}{c}{Composition / \si{\atpercent}} \\
	& state & & \multicolumn{1}{c}{Ti} & \multicolumn{1}{c}{Al} & \multicolumn{1}{c}{O} & \multicolumn{1}{c}{V} & \multicolumn{1}{c}{Mo} & \multicolumn{1}{c}{N} \\
	\hline
	Ti--7Al & IWQ & Bulk & 84.0 & 14.4 & 0.8 &  &  & 0.7 \\
	
    & 49 d & Bulk & 83.3 & 15.2 & 0.8 &  &  & 0.6 \\
	& & \textalpha{} & 84.2 & 14.2 & 0.9 &  &  & 0.7 \\
	& & \textalpha{}$_2$ & 74.6 & 24.6 & 0.3 &  &  & 0.7 \\
	
	& 120 d & Bulk & 82.9 & 15.4 & 1.0 &  &  & 0.7 \\
	& & \textalpha{} & 83.3 & 14.8 & 1.1 &  &  & 0.7 \\
	& & \textalpha{}$_2$ & 74.1 & 24.7 & 0.4 &  &  & 0.8 \\
	
	~--0.25O & 120 d & Bulk & 81.8 & 16.4 & 1.1 & & & 0.7 \\
	& & \textalpha{} & 83.7 & 14.4 & 1.3 &  &  & 0.6 \\
	& & \textalpha{}$_2$ & 72.7 & 25.7 & 0.9 &  &  & 0.6 \\
	
	~--1.1V--0.25O & 120 d & Bulk & 82.2 & 15.2 & 1.0 & 1.2 &  & 0.4 \\
	& & \textalpha{} & 83.6 & 13.7 & 1.1 & 1.3 &  & 0.4 \\
	& & \textalpha{}$_2$ & 72.4 & 25.5 & 0.6 & 0.9 &  & 0.6 \\
	
	~--0.8Mo--0.25O & 120 d & Bulk & 83.7 & 14.5 & 1.0 &  & 0.4 & 0.4 \\
	& & \textalpha{} & 85.2 & 12.9 & 1.1 &  & 0.4 & 0.4 \\
	& & \textalpha{}$_2$ & 75.8 & 22.7 & 0.6 &  & 0.4 & 0.4 \\
	\hline
	\end{tabular}
\end{small}\end{table}

\subsection{Small angle X-ray scattering}

\begin{figure*}[t!]
   \centering
   \includegraphics[width=135mm]{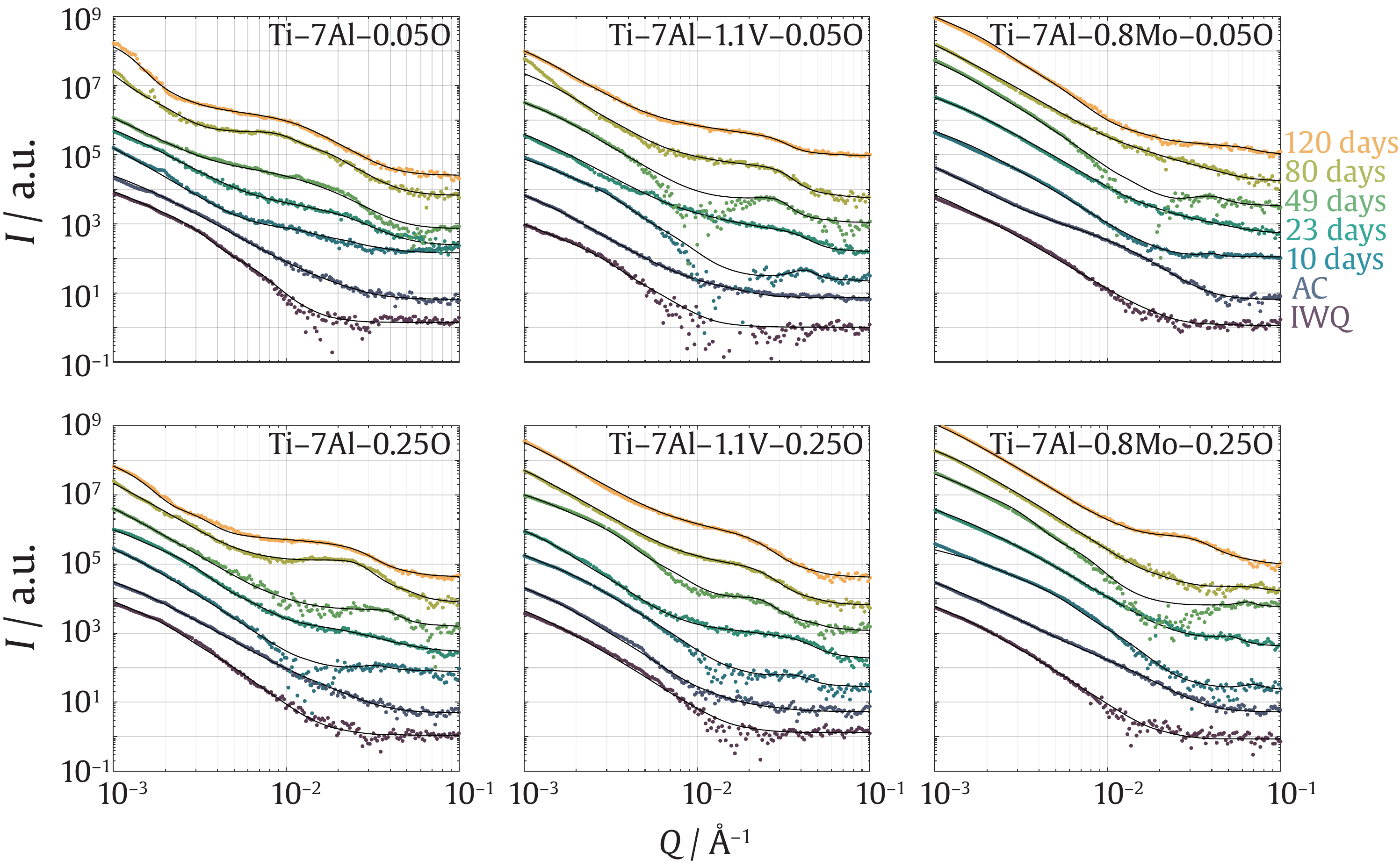}
   \caption{SAXS data obtained for the ageing study of the Ti--7Al model alloy series. In each case, a low-\textit{Q} peak is present due to scattering from grain boundaries, and a peak at high \textit{Q} develops with ageing time as the \textalpha{}$_2$ phase forms and grows. The variable influence of structure factor effects can be seen in different specimens, e.g. the Ti--7Al--1.1V--0.05O 10-day sample, as a dip in intensity at the low-\textit{Q} shoulder of the \textalpha{}$_2$ peak.}
   \label{fig:saxstime}
\end{figure*}

SAXS curves for each alloy in the ageing study, Fig.~\ref{fig:saxstime}, showed evolution of a peak at high scattering vector (\textit{Q}) over time, along with a low-\textit{Q} peak that showed no systematic variation with the ageing process. These are ascribed to \textalpha{}$_2$ precipitation and to the presence of grain boundaries as large scatterers respectively. The \textalpha{}$_2$ peak is distinctly visible for larger precipitates, but for very fine dispersions as in the Mo-containing materials it is less easily discerned. For all alloys, a slight difference in curve shape at high \textit{Q} between quenched and air-cooled states is seen.

The raw data were fitted using two scatterer populations for each of the main features. A prolate spheroidal model was used for \textalpha{}$_2$ precipitates, based on TEM images showing that precipitate aspect ratio increases with time. This shape is described with a shorter equatorial radius $r_e$ and longer polar radius $r_p$, so that particle aspect ratio is given by $A = r_p/r_e$ and precipitate volume can be calculated as $V = \frac{4}{3}\pi{}r_{e}^{3}A$. As seen in Fig.~\ref{fig:saxstime}, a structure factor effect occurs in some of the samples (appearing as a dip in measured intensity at the low-\textit{Q} shoulder of the \textalpha{}$_2$ peak). This effect is strengthened or subdued according to the competing effects of precipitate growth and coarsening on the extent to which each dispersion can be considered dilute.

Contrast values were calculated using the Irena analysis package \cite{Ilavsky2009}, using composition data from APT. The \textalpha{} and \textalpha{}$_2$ compositions for each sample were used to calculate the average atomic weight of each phase. This was then used to estimate the density of the phase, assuming no difference in unit cell volume compared to pure Ti, and these phase compositions and densities were then used to calculate scattering length density contrast. After calculating this for the different samples, values between 1.7--\SI{2.7E20}{\centi\metre^{-4}} were obtained but showed no systematic variation with alloy composition, so an average value of \SI{2.2E20}{\centi\metre^{-4}} was taken for fitting of all SAXS datasets in this study.

In the fitting process, $A$ is an input parameter along with phase contrasts calculated in the Irena analysis suite using APT data. Model outputs include $r_e$ and volume fraction of \textalpha{}$_2$ phase, $f$. These can be used to calculate precipitate number density, $n = f/v$, and if, in the absence of any specific model or description, a simple cubic array of precipitates in the matrix is assumed, the average spacing may be calculated as $s_\mathit{eff} = n^{-1/3}$. The low-\textit{Q} peak associated with grain boundaries was modelled as a cylindrical disc of appropriate diameter and thickness, and fitted using an arbitrary contrast that was not deconvolved from volume fraction for this microstructural component.

Direct comparison of $f$ between alloys is possible but, due to the variation in precipitate aspect ratio with both alloy and ageing time, the model output parameter $r_e$ is not an appropriate metric for comparing precipitate sizes across the study. A directly comparable quantity is the average volume of a precipitate, $V = \frac{4}{3}\pi{}r_{e}^{3}A$. For easy cross-reference with TEM images and APT reconstructions, the diameter of a sphere having equal volume to the modelled spheroid can be calculated as $d_{eq} = 2r_eA^{1/3}$. This equivalent sphere diameter provides a directly comparable metric of precipitate size at different ageing times, and allows easy comparison with TEM and APT data.

\begin{figure}[t!]
   \centering
   \includegraphics[width=90mm]{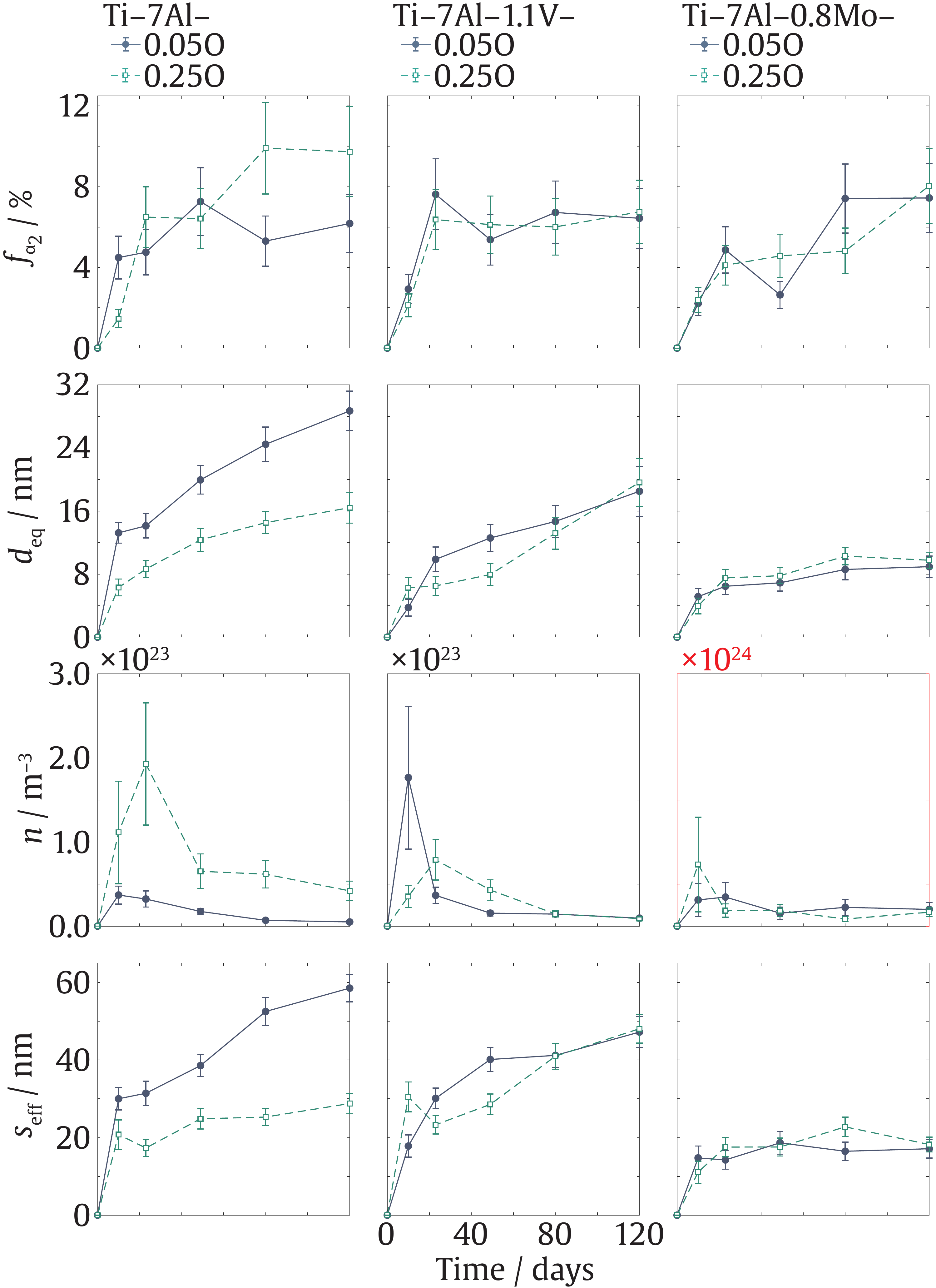}
   \caption{SAXS-derived quantitative analysis of the evolution of volume fraction $f_{\alpha_2}$, precipitate equivalent diameter $d_{eq}$, number density $n$ and effective square spacing $s_\mathit{eff}$. N.B. the $10\times$ change in scale for the number density in the Mo-containing samples (shown with red axes); this solute has the most significant effect on \textit{n}.}
   \label{fig:saxsoutcome}
\end{figure}

Fitting results for \textalpha{}$_2$ volume fraction, size and spacing showed similar trends with time for each alloy composition, but clear differences were apparent between different alloys, Fig.~\ref{fig:saxsoutcome}. Volume fraction $f$ showed the expected rapid initial increase followed by plateauing towards the equilibrium volume fraction for each system. An equilibrium volume fraction was reached for the base alloy and variants containing additions of O, V or both. A higher fraction of \textalpha{}$_2$ was observed in Ti--7Al--0.25O compared to Ti--7Al. The addition of V alone caused no significant difference in volume fraction, and the addition of O to the V-containing alloys did not influence the fraction of \textalpha{}$_2$. The Mo-containing alloys did not appear to reach an equilibrium state, with volume fraction still apparently increasing after 120 days at \SI{550}{\celsius}. The final measured volume fractions for these alloys after 120 days were slightly higher than for the base alloy. As with the V-containing alloys, no significant difference in volume fraction was seen between the Mo alloys with different O levels.

Precipitate size and spacing were reduced compared to the Ti--7Al base alloy upon adding any of the three solutes investigated. Mo had the most significant effect, producing very fine dispersions of small, closely-spaced precipitates. As with volume fraction, precipitate size and spacing showed no significant influence from oxygen content in the V- and Mo-containing materials. In each system, number density was found to vary as expected for precipitate coarsening behaviour, with an initial rapid increase corresponding to nucleation and early growth of precipitates. This was followed by a more gradual decrease as the microstructure underwent coarsening, with larger precipitates growing at the expense of smaller ones.

Comparing IWQ and AC datasets (i.e. quenched/disorded and SRO/early nucleation stage), a difference in curve shape is consistently seen across the different alloys. This takes the form of an increased intensity across a broad \textit{Q} range from around 0.01 to \SI{0.05}{\per\angstrom}. Modelling of the AC datasets was attempted using a low aspect ratio spheroid, but produced unphysical modelling results.

\subsection{APT crystallography}
\begin{figure}[h]
   \centering
   \includegraphics[width=80mm]{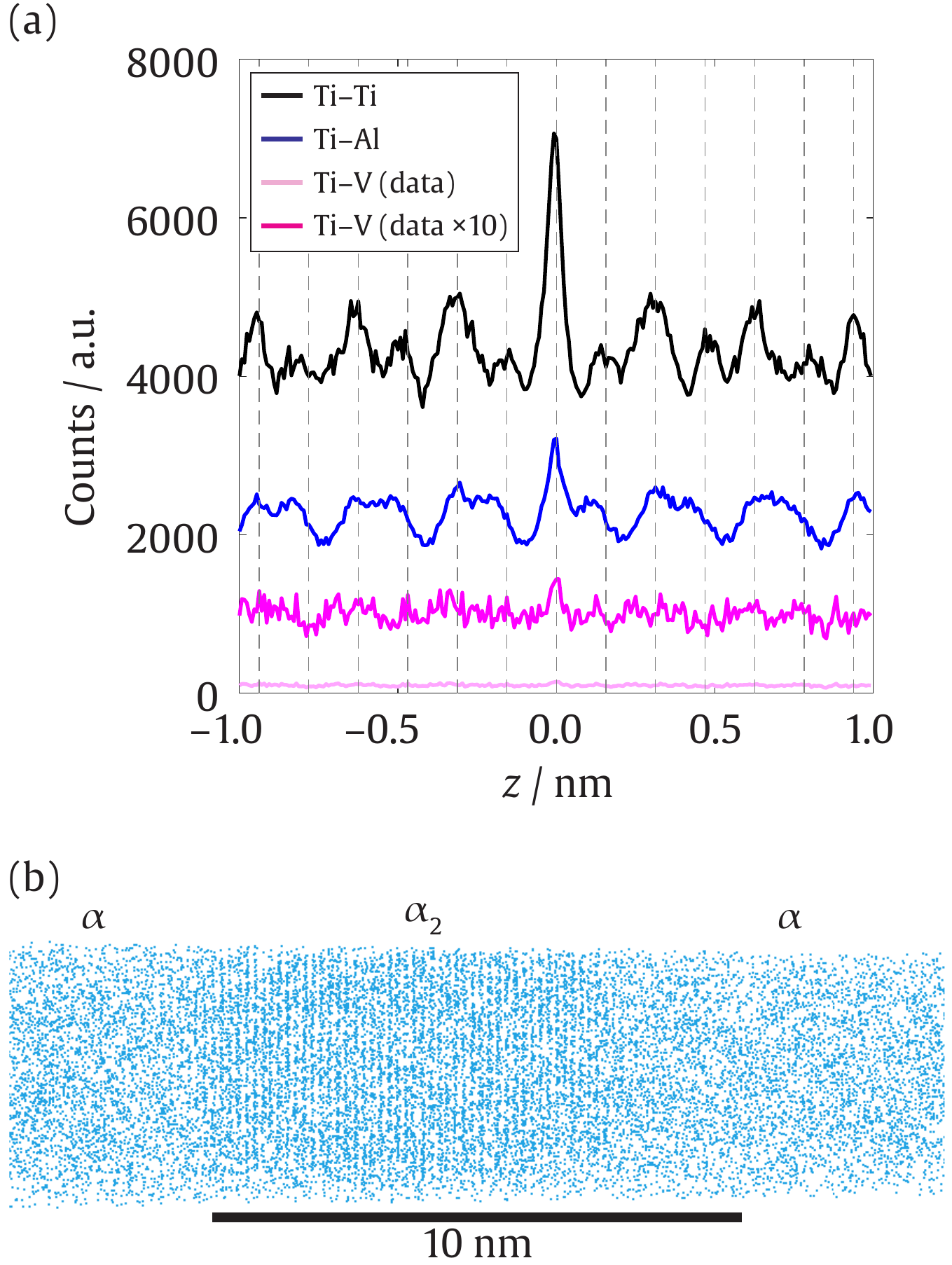}
   \caption{Crystallographic analysis was attempted for on-zone APT datasets containing ordered \textalpha{}$_2$ Ti$_3$Al precipitates, shown here for TI--7Al--1.1V--0.25O (wt.\%). For samples analysed parallel to a \hkl<2 -1 -1 0> direction, \textalpha{}$_2$ precipitates lying on a crystallographic pole in the reconstruction were analysed to produce spatial distribution maps, (a). An artefact previously described by Vurpillot \textit{et al.} \cite{Vurpillot2000} was seen in the interplanar spacings analysed for this dataset in both the spatial distribution maps (a) and the Al atom maps (b).}
   \label{fig:aptx}
\end{figure}

Attempts were made to analyse the APT data for crystallographic information, specifically regarding site partitioning of substitutional solutes V and Mo on the \textalpha{}$_2$ lattice. Spatial distribution maps (SDMs) were calculated for individual precipitates that had been identified as being located directly on a \hkl{2 -1 -1 0} pole, for Ti--Ti, Ti--Al and Ti--V species pairs, Fig.~\ref{fig:aptx}. However, due to the significant differences in evaporation field between Ti and Al under the measurement conditions applicable to these alloys, artefacts were seen in the interplanar spacing both in atom maps and in SDMs. This artefact has been described by Vurpillot \textit{et al.} \cite{Vurpillot2000}. By comparison, in Ni--Al \textgamma{}--\textgamma{}$^\prime$ alloys, the evaporation field difference is smaller so that site partitioning is more easily accessible through on-zone APT \cite{Bagot2017}. The low solubility of V and Mo in the \textalpha{} and \textalpha{}$_2$ phases also made this analysis challenging due to limited V or Mo atoms available for measurement and SDM analysis.

\section{Discussion}

The use of TEM, APT and SAXS in combination has allowed the analysis of volume fraction, size and spacing of \textalpha{}$_2$ in this set of model alloys. Comparisons between the various alloying elements may be made.

\subsection{Volume fractions}

First, considering volume fraction, the increase from 6\% to 10\% due to increased oxygen content supports the small shift in the position of the \textalpha{}/\textalpha{}+\textalpha{}$_2$ boundary upon adding oxygen that has been previously suggested \cite{Lim1976,Waterstrat1988,Gray1990}. Based on the results of this study, vanadium does not appear to significantly alter the volume fraction of \textalpha{}$_2$ produced after 120~days ageing at \SI{550}{\celsius}. Molybdenum causes a slight increase from 6\% to 8\%, but due to retardation of phase separation kinetics by this solute, the Mo-containing systems did not reach equilibrium volume fractions during the 120~days of this study.

\subsection{Size and spacing}

Regarding the size and spacing of precipitates, these were largest at all times for the base alloy Ti--7Al--0.05O. Additions of any of the three solutes investigated caused refinement of the \textalpha{}$_2$ dispersion. Molybdenum had the most significant effect on this, followed by oxygen, while vanadium had a fairly minimal effect on the size and spacing of precipitates. The varying degrees of refinement are reflected in the extent to which each sample's scattering curve shows structure factor effects. Considering also the number density at short and long times for the different alloys, all solutes are seen to increase \textit{n} during the early stages of phase separation. This suggests that adding any of these solutes causes increased nucleation density. Molybdenum produces an order of magnitude increase in early number density compared to the other alloys in the study. It is suggested that the resulting reduction in interparticle distances then causes smaller precipitate sizes due to soft impingement.

In the V- and Mo-containing alloys, there is little difference in volume fraction, size and spacing of \textalpha{}$_2$ precipitates between the low- and high-oxygen variants in each case. This indicates that the \textalpha{}/\textalpha{}+\textalpha{}$_2$ boundary in the Ti--Al--O--V and Ti--Al--O--Mo quaternary systems becomes less sensitive to O as the \textbeta{} stabiliser content increases.

\subsection{Coarsening and LSW modelling}

In order to compare the precipitate growth rate and coarsening between alloys, the Lifshitz--Slyozov--Wagner (LSW) model was applied for the precipitate effective radius, \textit{r} = \textit{d}$_{\mathrm{eq}}/2$ (for direct comparison between samples with different precipitate aspect ratios). This model describes the evolution of precipitate size with time according to
\[r^{3}(t)-r_{0}^{3} = \frac{8\mathit{\Gamma{}}DCV_{m}^{2}}{9RT}t = K_{\mathrm{LSW}}t,\]

\noindent where $\mathit{\Gamma{}}$ is the precipitate/matrix interfacial energy, $D = D_{0}exp(-Q/RT)$ is the diffusion coefficient of the rate-limiting species through the matrix, $C$ is the equilibrium concentration of the rate-limiting species in the matrix, $V_{m}$ is the molar volume of the precipitate phase, $R$ is the ideal gas constant and $T$ is the absolute temperature at which phase separation has been observed \cite{Lifshitz1961,Wagner1961}. The gradients of linear fits hence provide a rate constant, $K_{\mathrm{LSW}}$, that can be compared between alloys. There may to be overlap between the growth and coarsening regimes for the \textalpha{}$_2$ dispersions observed, Fig.~\ref{fig:saxsoutcome}; in the first few days and in the Mo-containing alloys the phase fraction increases with time. The LSW approach does not deconvolve these two processes, such that the extracted $K_{\mathrm{LSW}}$ are not pure coarsening rates, especially in the case of Mo, but also incorporate growth rate to an extent depending on the degree of overlap between growth and coarsening stages. Nonetheless, this analysis allows some comparison of the phase separation kinetics between alloys. It is interesting to note that in the Ti--7Al--0.25O and V-containing alloys, a fairly consistent coarsening rate of 4--\SI{7}{\nano\meter\cubed\per\day} is obtained.

\begin{figure}[t!]
   \centering
   \includegraphics[width=88mm]{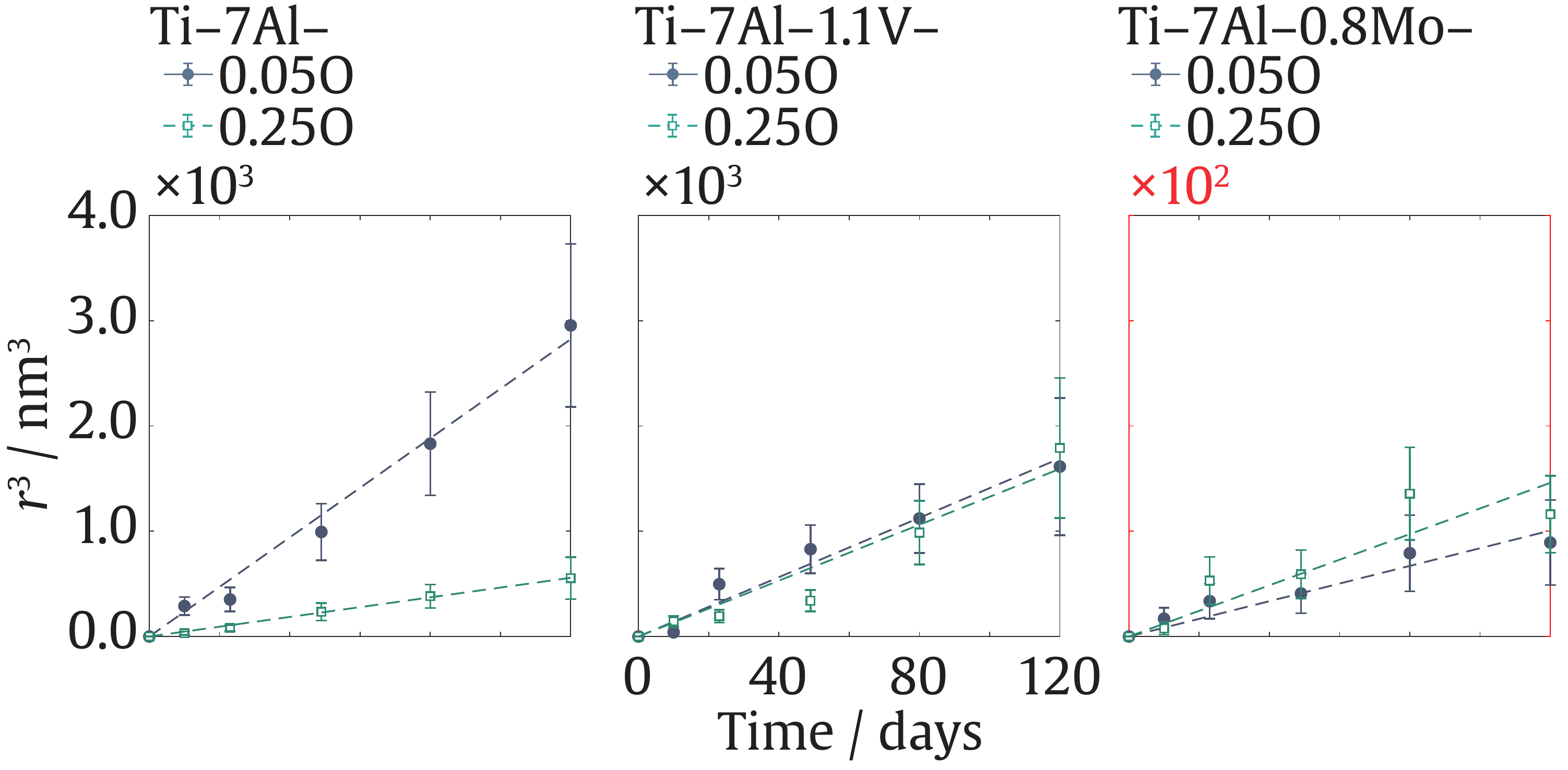}
   \caption{Lifshitz--Slyozov--Wagner modelling assumes that a precipitation coarsening process is controlled by the diffusion of a rate-limiting species through the matrix, leading to a linear proportionality between precipitate volume and time. Plotting $r^{3}$ against time, good linear fits were obtained for each alloy in the study, supporting a matrix diffusion-controlled coarsening mechanism for \textalpha{}$_2$ in \textalpha{}-Ti--Al.}
   \label{fig:lsw}
\end{figure}

\begin{table}[t!!]\begin{small}
	\centering \setlength{\tabcolsep}{2pt}
	\caption{Lifshitz--Slyozov--Wagner modelling was successfully applied to the SAXS data, Fig.~\ref{fig:lsw}, with least-squares goodness of fit \textit{R}$_{LS}^{2}$ of 0.91 or better. This model provides coarsening rate constants $K_{\mathrm{LSW}}$ and an estimate of the coarsening rate of precipitates in each alloy in terms of volume per unit time. It is noted that, for these alloys, there is likely a significant overlap of the growth and coarsening regimes. The peak precipitate number density during the coarsening process, as analysed in SAXS, is shown for comparison.}
	\begin{tabular}{l|c c c|c}
	\hline
	Material & $K_{\mathrm{LSW}}$ & \textit{R}$_{LS}^{2}$ & Coarsening rate & Peak $n$\\
	 & / \SI{E-31}{\meter\cubed\per\second} & & / \si{\nano\meter\cubed\per\day}\ & / 10$^{22}$~m$^{-3}$\\
	\hline
	Ti--7Al & 1.7$\pm$0.1 & 0.98 & 15$\pm$1 & 3.7 \\
	~--0.25O & 0.46$\pm$0.03 & 0.98 & 4.0$\pm$0.3 & 19.3 \\
	~--1.1V & 0.83$\pm$0.07 & 0.97 & 7.2$\pm$0.6 & 7.9 \\
	~--1.1V--0.25O & 0.79$\pm$0.07 & 0.96 & 6.8$\pm$0.6 & 17.7 \\
	~--0.8Mo & 0.08$\pm$0.01 & 0.96 & 0.68$\pm$0.06 & 34.6 \\
	~--0.8Mo--0.25O & 0.13$\pm$0.02 & 0.91 & 1.1$\pm$0.2 & 73.3 \\
	\hline
	\end{tabular}
	\label{table:lsw}
\end{small}\end{table}

A nonzero value of $r_0$ would indicate an incubation period between the introduction of isothermal ageing conditions and the onset of precipitation. In this study, $r_{0} = 0$ was set for LSW fitting as no incubation period is expected nor evident in the data. Least squares linear regression for each alloy gave goodness of fit \textit{R}$_{LS}^{2}$ values of 0.9 or above for a \nicefrac{1}{3} power law, Fig.~\ref{fig:lsw}, indicating a diffusion-limited growth process, rather than an interface-controlled mechanism \cite{Ardell2013}. The $K_{\mathrm{LSW}}$ values obtained for these alloys, Table~3, are reasonable when compared to those reported in an analagous study of \textgamma{}--\textgamma{}$^{\prime}$ Ni superalloys \cite{vorontsov2016}, considering the much slower formation of \textalpha{}$_2$ in Ti--Al alloys than of \textgamma{}$^{\prime}$ in Ni--Al alloys.

Comparing the rates obtained for the different alloy compositions, Table~3, and considering that a matrix diffusion-controlled growth mechanism is supported by the good fits obtained for the LSW model, it may be anticipated that the growth rate would depend only on the diffusivity of the key species, aluminium. Although Mo has been seen to slow phase separation kinetics for the \textalpha{}$\rightarrow$\textbeta{} transformation \cite{Ackerman2020}, in this instance the Mo concentration is very low and considered unlikely to have such a stark effect on number density through modification of Al diffusivity alone. It is suggested instead that, due to soft impingement of solute fields around growing \textalpha{}$_2$ precipitates, the coarsening rate is controlled by the nucleation density of the precipitate dispersion, reflected by a correspondence between rate and interprecipitate spacing $s_{\mathrm{eff}}$.

A further outcome of this study is that a value for the \textalpha{}/\textalpha{}$_2$ interfacial energy may be calculated, since this is a term contained in the rate constant $K_{\mathrm{LSW}}$. Using the values given in Table~\ref{table:lswdata}, a value of $\mathit{\Gamma{}} =$~107~mJ~m$^{-2}$ was obtained for the base alloy Ti--7Al--0.05O. Since coarsening of Ti$_3$Al in Ti has not been quantified before, literature estimates for comparison are not available. It should be noted that the value inferred for $\mathit{\Gamma{}}$ depends strongly on that assumed for $D$, which is itself an extrapolation from literature measurements at higher temperatures, and may be strongly affected by e.g. vacancy and minor solute content in the samples studied. For a factor of 10 change to the coarsening rate, if assumed to be entirely driven by $\mathit{\Gamma{}}$, this would imply an order of magnitude difference in the size of $\mathit{\Gamma{}}$ between the alloys in this series, since the relationship is linear. This is unlikely given only minor differences in composition between alloys; a more sophisticated coarsening and growth analysis would be required to fully deconvolve the effects for the Mo-containing alloys, rather than the simplest possible dilute coarsening LSW analysis performed here.

\begin{table}[h!]\begin{small}
	\centering
	\caption{Parameters used in the calculation to estimate \textalpha{}/\textalpha{}$_2$ interfacial energy based on LSW fitting for the Ti--7Al--0.05O alloy. The equilibrium concentration $C_{\mathrm{Al}}$ corresponds to the \SI{14.8}{\atpercent}~Al measured for the \textalpha{} phase of this alloy after 120~d ageing at \SI{550}{\celsius}. $D_{\mathrm{Al}}(550~^{\circ{}}\mathrm{C})$ and $V_{\mathrm{m}}$ are calculated using data from \cite{LandW}.}
	\begin{tabular}{c c c c}
	\hline
	$K_{\mathrm{LSW}}$ & $D_{\mathrm{Al}}(550~^{\circ{}}\mathrm{C})$ & $C_{\mathrm{Al}}$ & $V_{\mathrm{m}}$\\
	 / \SI{E-31}{\meter\cubed\per\second} & / \si{\metre\squared\per\second} & / \si{\mol\per\metre\cubed} & / \si{\metre\cubed\per\mol}\\
	\hline
	1.7 & $6\times{}10^{-24}$ & $12.1\times{}10^{3}$ & $10.7\times{}10^{-6}$\\
	\hline
	\label{table:lswdata}
	\end{tabular}
\end{small}\end{table}

\subsection{Effect of quenching temperature on \textalpha{}$_2$ formation}

A second heat treatment study was conducted in order to establish a clearer mechanistic connection between the tertiary solutes (O, V, Mo) and differences in \textalpha{}$_2$ formation. Upon adding any of the three solutes, refinement of the \textalpha{}$_2$ dispersion was observed, to a greater or lesser extent depending on the solutes included. Noting the homogeneous distribution of \textalpha{}$_2$ precipitates across \textalpha{} grains in these alloys, it was suggested that the nucleation points must correspond to a homogeneously distributed lattice defect. A plausible candidate is the vacancy concentration in each alloy. It was proposed that, upon adding any solute, the resulting increased entropy of the alloy causes an increase in vacancy concentration, and that this is the common underlying mechanism controlling nucleation density.

\begin{figure}[b!]
   \centering
   \includegraphics[width=80mm]{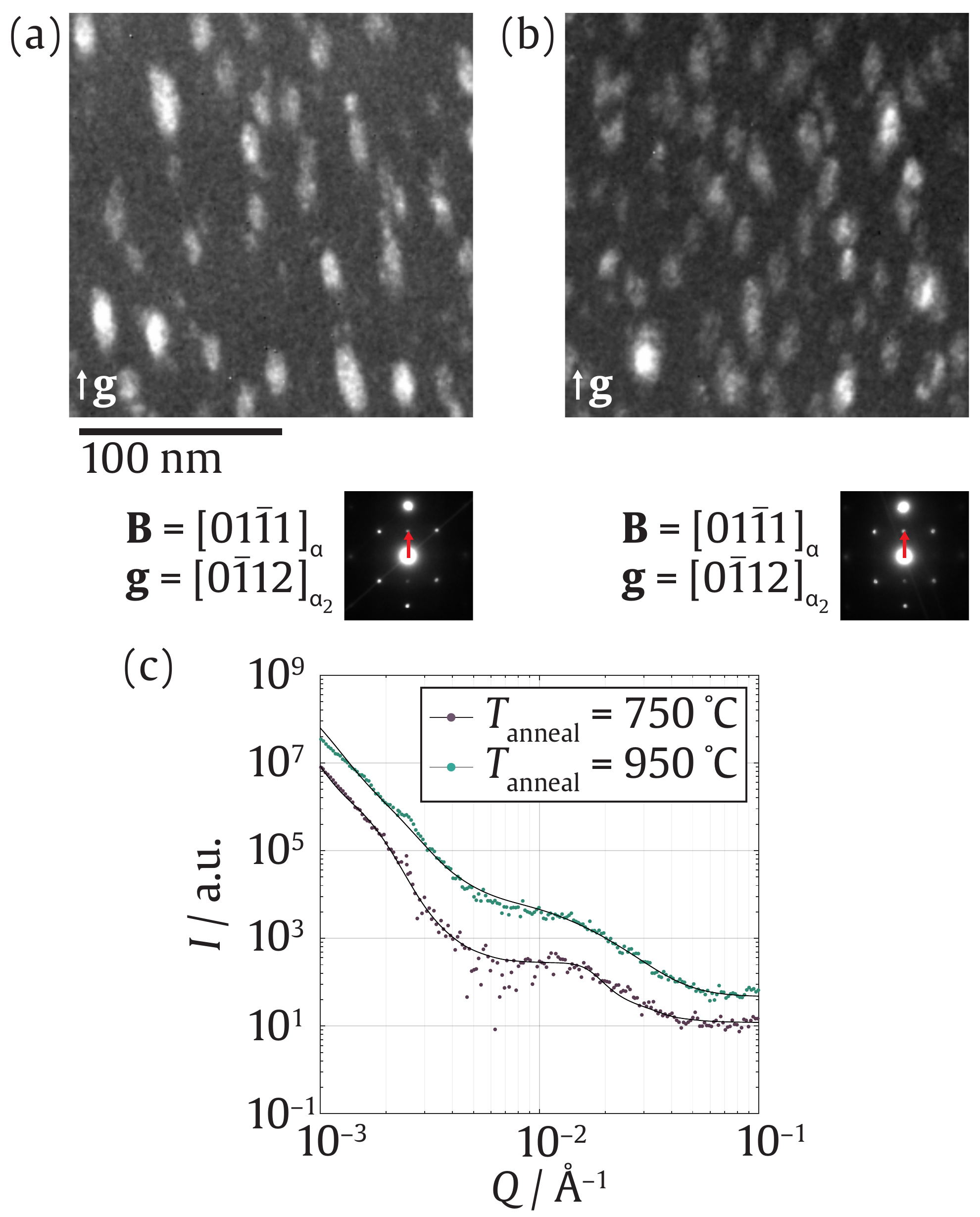}
   \caption{Effect of quenching temperature on $\alpha_2$ formation, where Ti--7Al--0.05O specimens were aged for 23~days after quenching from annealing temperatures $T_{\mathrm{anneal}}$ (a) \SI{750}{\celsius} and (b) \SI{950}{\celsius}, freezing in lower and higher vacancy concen trations prior to ageing respectively. Dark field TEM shows the expected spheroidal morphology and a qualitative indication of different dispersion characteristics. SAXS measurements of these samples showed clear differences in the position and intensity of the \textalpha{}$_2$ peak, as well as the influence of structure factor (associated with higher number density) for the \SI{950}{\celsius} sample.}
   \label{fig:vacresults}
\end{figure}

To establish whether a link exists between vacancy concentration and \textalpha{}$_2$ nucleation density, a single alloy composition was used, Ti--7Al--0.05O, while the thermal history of the samples prior to \textalpha{}$_2$ ageing was varied. Vacancy concentration in metals is known to have an empirically Arrhenius-type dependence on temperature. Aiming to control the vacancy concentration in samples prior to ageing, pieces of the alloy in the IWQ starting condition were annealed at \SI{750}{\celsius} and \SI{950}{\celsius} to generate two different vacancy concentrations while remaining within the \textalpha{} phase field and staying above \textalpha{}$_2$ formation temperatures. The samples were then ice water quenched, cleaned to remove any oxide, and encapsulated under argon in a quartz ampoule before ageing at \SI{550}{\celsius} for 23~days. The resulting \textalpha{}$_2$ dispersions were then characterised using dark field TEM imaging and SAXS measurements Fig.~\ref{fig:vacresults}. The SAXS data were fitted using a spheroidal precipitate shape with an aspect ratio of 2.0 and a contrast value of 2.2~$\times{}~10^{-20}$~cm$^{-4}$, following the same methodology as for the main SAXS dataset.

In order to establish whether these results are consistent with a vacancy-controlled nucleation mechanism, the number densities in each sample were compared to predicted vacancy concentration behaviour. The vacancy concentration at a temperature $T$ is empirically given by
\[
N = N_{0}e^{-E_{\mathrm{f}}/k_{\mathrm{B}}T},\]
where $N_{0}$ is a constant prefactor and $E_{\mathrm{f}}$ is the vacancy formation energy. For two different temperatures, $T_1$ and $T_2$,
\[
\frac{N_1}{N_2} = \frac{\mathrm{exp}(-E_{\mathrm{f}}/k_{\mathrm{B}}T)}{\mathrm{exp}(-E_{\mathrm{f}}/k_{\mathrm{B}}T)},\]
such that
\[
E_{\mathrm{f}} = k_{\mathrm{B}} \left(\frac{1}{T_{2}}-\frac{1}{T_{1}}\right)^{-1} \mathrm{ln}\left(\frac{N_{1}}{N_{2}}\right).\]
If it is assumed that $n \propto{} N_{0}$ independent of temperature, then the \textalpha{}$_2$ number densities $n$ may be used to estimate the vacancy formation energy in Ti--7Al--0.05O.

For the samples quenched from \SI{750}{\celsius} and \SI{950}{\celsius}, number densities of $2.2\times{}10^{21}$~m$^{-3}$ and $3.5\times{}10^{22}$~m$^{-3}$ were obtained from SAXS data fitting respectively, Table~\ref{table:vacancy}. This gives an estimate of $E_{\mathrm{f}} = 1.5\pm{}0.4$~eV. Previous experimental studies for vacancy formation in \textalpha{}-Ti are relatively scarce, but comparison may be made with the findings of Hashimoto~\textit{et~al.} \cite{Hashimoto1984} who measured a value of $1.27\pm{}0.05$~eV using positron annihilation. First-principles calculations for $E_{\mathrm{f}}$ in \textalpha{}-Ti have found values of 1.87~eV \cite{Connetable2011} and 1.97~eV \cite{Raji2009}. The value obtained in the present work is consistent with these earlier studies, lending support to a vacancy-mediated nucleation mechanism for \textalpha{}$_2$ in \textalpha{}-Ti--Al--X alloys, as a second-order effect of tertiary solute additions.

\begin{table}[h!]\begin{small}
    \centering
	\caption{SAXS fitting results for specimens solutionised at and quenched from different solutionising temperatures, \textit{T}$_{sol}$, prior to ageing for 23~days at \SI{550}{\celsius} in order to compare the effects of different vacancy concentrations on volume fraction \textit{f}$_{\alpha_{2}}$, precipitate size as equivalent sphere diameter \textit{d}$_{\mathrm{eq}}$ and precipitate number density $n$. For both materials, a contrast value of \SI{2.2E20}~cm$^{-4}$ and an aspect ratio of 2.0 were used.}
	\begin{tabular}{c c c c c c}
	\hline
	\textit{T}$_{sol}$ / \si{\celsius} & \textit{f}$_{\alpha_{2}}$ & \textit{d}$_{\mathrm{eq}}$ / nm & \textit{n} / $10^{21}$ m$^{-3}$\\
	\hline	
	750 & 0.012 & 21.8 & 2.21 \\
	950 & 0.058 & 14.7 & 34.9 \\
	\hline
    \label{table:vacancy}
    \end{tabular}
\end{small}\end{table}

\section{Conclusions}\label{conclusions}
In this study, the precipitation and coarsening of \textalpha{}$_2$ Ti$_3$Al in a Ti--Al alloy series was studied, using TEM to identify the existence of the precipitates and their morphology, APT to characterise their composition, and SAXS to quantify their number density, size and fraction. The effect of interstitial solute O and substitutional solute V and Mo, and of quenching temperature, were examined. The following conclusions are drawn.

\begin{itemize}\setlength{\itemsep}{-1mm}
\item Interstitial O increases the volume fraction of \textalpha{}$_2$ formed at equilibrium, which is in the region of 8-10 vol\%, while V and Mo have a relatively small effect. The precipitates grow in size to up to \SI{30}{\nano\metre} after \SI{120}{\day} at \SI{550}{\celsius} (Ti--7Al--0.05O), with interparticle spacing of a similar magnitude;
\item Addition of O, V or Mo increases the nucleation density of \textalpha{}$_2$, and leads to a finer precipitate dispersion;
\item Growth of \textalpha{}$_2$ can be described using an LSW model, indicating diffusion control (rather than interface coherency control);
\item A secondary study comparing \textalpha{}$_2$ formation between samples differing only by quenching temperature showed a difference in nucleation number density. This gave an activation energy consistent with a vacancy nucleation mechanism, $E_{\mathrm{f}} = 1.5\pm{}0.4$~eV;
\item This leads to the inference that the effect of solute O, V and Mo is, broadly, to increase the nucleation number density and thereby slow coarsening due to an earlier onset of soft impingement.
\end{itemize}

\section*{Acknowledgements}
\noindent\small FFD was funded by Rolls-Royce plc and by the EPSRC Centre for Doctoral Training in the Advanced Characterisation of Materials (EP/L015277/1). DD was funded by a Royal Society Indstrial Fellowship and EPSRC (EP/K034332/1). BG and PK are grateful for funding from the Max Planck Society through the Laplace project. The authors are grateful to U. Tezins, C. Bross and A. Sturm for their technical support of the APT and FIB facilities at the Max-Planck Institut f\"{u}r Eisenforschung, and for useful discussions with S. Balachandran. This research used resources of the Advanced Photon Source, a U.S. Department of Energy (DOE) Office of Science User Facility operated for the DOE Office of Science by Argonne National Laboratory under Contract No. DE-AC02-06CH11357. Useful conversations and technical assistance are also gratefully acknowledged from D. Isheim at Northwestern and A. Minor, R. Zhang and R. Traylor at UC Berkeley and Lawrence Berkeley National Laboratory, along with the help of K.M. Rahman and I. Bantounas at Imperial with the alloy processing.

\bibliographystyle{model1-num-names}
\bibliography{paper1.bib}

\end{document}